\documentclass[aps, prx,twocolumn,showpacs,floatfix,superscriptaddress,nofootinbib]{revtex4-2}
\usepackage{graphics}
\usepackage{xcolor}
\usepackage{amsmath, amssymb,amsthm}
\usepackage{mathtools}
\usepackage{braket}
\usepackage{bbold}
\usepackage[normalem]{ulem}
\usepackage[colorlinks=true,citecolor=blue,linkcolor=magenta]{hyperref}
\usepackage[caption=false]{subfig}

\newcommand{\diff}[1]{\mathrm{d}{#1}\,}

\definecolor{MyBlue}{HTML}{1f77b4}
\definecolor{MyOrange}{HTML}{ff7f0e}
\definecolor{MyGreen}{HTML}{2ca02c}
\definecolor{MyRed}{HTML}{d62728}

\newcommand{\msout}[1]{\text{\sout{\ensuremath{#1}}}}

\newcommand{\mold}[1]{\textcolot{blue}{\msout}}

\newcommand{\new}[1]{\textcolor{black}{#1}}

\begin{document}

\title{Many-body quantum boomerang effect}

\author{Jakub Janarek}\email{jakub.janarek@uj.edu.pl}
\affiliation{Laboratoire Kastler Brossel, Sorbonne Universit\'e, CNRS,
ENS-PSL Research University, Coll\`ege de France, 4 Place Jussieu, 75005
Paris, France}
\affiliation{Instytut Fizyki Teoretycznej, 
Uniwersytet Jagiello\'nski,  \L{}ojasiewicza 11, PL-30-348 Krak\'ow, Poland}

\author{Jakub Zakrzewski}\email{jakub.zakrzewski@uj.edu.pl}
\affiliation{Instytut Fizyki Teoretycznej,
Uniwersytet Jagiello\'nski,  \L{}ojasiewicza 11, PL-30-348 Krak\'ow, Poland}
\affiliation{Mark Kac Complex Systems Research Center, Uniwersytet Jagiello{\'n}ski, Krak{\'o}w, Poland}

\author{Dominique Delande}\email{dominique.delande@lkb.upmc.fr}
\affiliation{Laboratoire Kastler Brossel, Sorbonne Universit\'e, CNRS,
ENS-PSL Research University, Coll\`ege de France, 4 Place Jussieu, 75005
Paris, France}

\date{\today}

\begin{abstract}
We study numerically the impact of many-body interactions on the quantum boomerang effect. We consider {various} cases:  weakly interacting bosons, {the} Tonks-Girardeau gas, and strongly interacting bosons (which may be mapped {on}to weakly interacting fermions). Numerical simulations are performed using {the} time-evolving block decimation algorithm{, a quasi-exact method based on matrix product states}. In the case of weakly interacting bosons, we find a partial destruction of the quantum boomerang effect, in agreement with the earlier  mean-field study [Phys. Rev. A \textbf{102}, 013303 (2020)]. For the Tonks-Girardeau gas, we show the presence of the full quantum boomerang effect. For {strongly interacting bosons,} we observe a partial boomerang effect. We show that the destruction of the quantum boomerang effect is universal and does not depend on the details of the interaction between particles.
\end{abstract}
\date{\today}

\maketitle

\section{Introduction}
Anderson localization (AL), i.e. inhibition of transport, is one of the most famous phenomena in disordered systems \cite{Anderson1958}. It was successfully observed in many experiments, including quantum systems  \cite{Roati2008, Billy2008, Chabe2008, Jendrzejewski2012, Manai2015, Semeghini2015,White2020}, light \cite{Chabanov2000, Schwartz2007}, and sound waves \cite{Hu2008} among many others. Despite the many years since the first work on AL, a new phenomenon has recently been discovered that is a direct manifestation of localization: the quantum boomerang effect (QBE) \cite{Prat2019}. The new phenomenon involves the dynamics of wave packets with non-zero initial velocity evolving in Anderson localized systems. \new{Being related to Anderson localization, the boomerang effect should
exist for any type of waves exhibiting Anderson localization.
In the following, for concreteness, we study the QBE for the
Schr\"{o}dinger equation.}

In an Anderson localized system, as shown in \cite{Prat2019}, the center of mass (CM) of a quantum wave packet with an initial velocity, on average, returns to its initial position. This behavior is very different from that observed for the classical counterpart: a classical particle will randomize its velocity and, on average, localize after traveling a finite distance (a transport mean free path). The QBE is a genuine quantum phenomenon occurring in one and higher-dimensional Anderson localized systems \cite{Prat2019}, as well as generalized systems including {the} kicked rotor \cite{Tessieri2021}, systems without time-reversal symmetry \cite{Janarek2022}, see also \cite{Macri22}, and non-Hermitian systems \cite{Noronha22}. Recently, the QBE was observed in a quantum kicked rotor experiment \cite{Sajjad2021}, where the U-turn of the average momentum was reported.

Consider an Anderson localized one-dimensional system with {the} Hamiltonian $H = p^2/2m + V(x)$, where $V(x)$ is a {disordered potential} \cite{Prat2019}.  We define the average CM as 
\begin{equation} 
\langle x (t)\rangle = \new{\overline{\int x{|\psi(x,t)|^2}\diff{x}}} = \int x\overline{|\psi(x,t)|^2}\diff{x},
\end{equation}
 where $\overline{(\ldots)}$ denotes averaging over disorder realizations.  The full quantum boomerang effect occurs iff the CM of a wave packet with a non-zero initial velocity (e.g. $\psi_0(x) = \mathcal{N}\exp(-x^2/2\sigma^2 + ik_0 x)$)  returns to its initial position, $\langle x(t=\infty)\rangle = 0$ for large times $t\rightarrow\infty$.

When interactions are present in the system, this behavior changes as interactions tend to weaken localization phenomena. In effect a full localization may be replaced by a subdiffusive evolution at long times in the presence of interactions \cite{Pikovsky2008, Skokos2009, Flach2009, Cherroret2014a}. In a previous study \cite{Janarek2020a} we have shown that the interactions treated within the mean-field approach using {the} Gross-Pitaevskii equation (GPE) \cite{Pitaevskii2016}, lead to a partial destruction of the QBE. After the initial evolution, typical for the full QBE, the CM {performs a U-turn but} does not fully return to its origin{, saturating}  at some finite value. This final CM position {depends on} the interaction strength via  the interaction energy. Moreover, it was shown that the destruction of the QBE may be described using a single characteristic time scale, dubbed the \emph{break time}, 
beyond which the QBE is destroyed by interactions \cite{Janarek2020a}.

In the present work, we investigate many-body interactions between particles using quasi-exact numerical methods. 
In the first part, we analyze weakly interacting bosons with contact interaction and compare their dynamics to the mean-field approximation results. The many-body interactions lead to a stronger destruction of the QBE. However, it is shown that the effective break time analysis is still valid in the many-body system. In the second part, we show the full QBE for {the} Tonks-Girardeau gas where we also present a full localization of the final particle density.
In the last part, we study strongly interacting bosons, which map to weakly interacting fermions with momentum-dependent interactions. Similarly to weakly interacting bosons, we observe a partial QBE only. 
We show that, also in this case, the destruction of the QBE can be captured using similar methods. The results presented reveal that the destruction of the full QBE does not depend on the details of the interaction between the particles. 

The paper is organized as follows. Section~\ref{sec:model} introduces the model and explains the method used for numerical simulations of the system. It also presents the main parameters of the system. In Sec.~\ref{sec:bosons}, we study the case of weakly interacting bosons, where we also present a comparison with the mean-field model. Section~\ref{sec:tg} presents the observation of the QBE for {the} Tonks-Girardeau gas, while in Sec.~\ref{sec:fermions} we study strongly interacting bosons and analyze results from the perspective of weakly interacting fermions. Finally, we conclude the paper in Sec.~\ref{sec:conclusions}.

\section{The model}\label{sec:model}
We study a one-dimensional many-body bosonic Hamiltonian:
\begin{equation}
    \label{eq:many_body_hamiltonian}
    \hat{H} =\! \int \!\hat{\Psi}^\dagger(x)\!\left( - \frac{\hbar^2}{2m}\Delta + V(x) + \frac{U}{2}\hat{\Psi}^\dagger(x) \hat{\Psi}(x) \right)\! \hat{\Psi}(x)\diff{x},
\end{equation}
where $m$ is the particle mass, $V(x)$ represents the disordered potential, and $U$ is the strength of the two-body contact potential. In our work, we adopt the method introduced in \cite{Schmidt2007} and map {the} continuous Hamiltonian (\ref{eq:many_body_hamiltonian}) to a discrete model on a\new{n} equidistant grid with $L$ lattice sites, where the position is given by $x_j = {j \Delta x}$, $j\in \mathbb{Z}$, and $\Delta x$ is {the} grid spacing. We start by expanding the field operators in the basis of bosonic annihilation operators $\hat{a}_j$ and single-particle wave functions $\psi_j(x)$: step functions localized at position $x_j$. {The field} operators decomposition is given by:
\begin{equation}
    \hat{\Psi}(x) = \sum_{j=1}^L \psi_j(x)\hat{a}_j.
\end{equation}
The derivative in Hamiltonian~(\ref{eq:many_body_hamiltonian}) is expressed as the three-point stencil, $\partial^2_x\hat{\Psi}(x_j) \to (\hat{\Psi}(x_{j-1}) - 2 \hat{\Psi}(x_j) + \hat{\Psi}(x_{j+1}))/\Delta x^2$. The resulting Hamiltonian has the form of {a disordered} Bose-Hubbard Hamiltonian:
\begin{equation}
    \label{eq:bose_hubbard}
    \hat{H} = -J_0\sum_{j=1}^L\left(\hat{a}_j^\dagger \hat{a}_j + \text{c.c.}\right) + \sum_{j=1}^L V_j \hat{n}_j + \frac{U_0}{2}\sum_{j=1}^L \hat{n}_j(\hat{n}_j-1),
\end{equation}
where the parameters are directly connected with the lattice spacing $\Delta x$ and the parameters of Hamiltonian (\ref{eq:many_body_hamiltonian}): $J_0 = \hbar^2/(2m\Delta x^2)$, $U_0 = U/\Delta x$, and $V_j = V(x_j)$. Thanks to this discretization technique, we are able to study the many-body QBE in a continuous space with techniques developed for lattice models.

In our work, we are almost exclusively interested in {the temporal dynamics} of the system. To compute the time evolution {under} the Bose-Hubbard Hamiltonian (\ref{eq:bose_hubbard}), we use a homemade implementation of the time-evolving block decimation (TEBD) algorithm \cite{Vidal2003, Vidal2004} based on matrix product states (MPS). At each time step, the many-body state is expressed in terms of matrices $\Gamma^{i_l}$ and vectors $\lambda^{[l]}$:
\begin{equation}\label{eq:psi_mps}
  \ket{\Psi} = \sum_{\substack{\alpha_1, \ldots,\alpha_L\\ i_1,\ldots,i_L}} \Gamma^{ i_1}_{1,\alpha_1} \lambda^{[1]}_{\alpha_1}\Gamma^{i_2}_{\alpha_1,\alpha_2} \new{\lambda^{[2]}_{\alpha_2}}\ldots\Gamma^{i_L}_{\alpha_L,1} \ket{i_1,i_2,\ldots,i_L},
\end{equation}
where matrices $\Gamma^{i_l}$ describe the $l$-th site and vectors $\lambda^{[l]}$ describe bonds between $l$ and $l+1$ sites. The indices $i_l$ run form 0 to maximal occupation $i_\text{max}$; the indices $\alpha_l$ run from 1 to $\chi$ (the so called bond dimension).

The QBE is a phenomenon {displayed} by wave packets having nonzero initial velocities. For  {a} single-particle wave function, a nonzero initial velocity translates into a nonzero phase factor $e^{ik_0x}$ {for the} wave function, where $k_0$ {is related to} the velocity by $v_0=\hbar k_0/m$. It means that it is possible to \emph{kick} a wave packet by multiplying its wave function by {a proper} phase factor. 

When the state is in the MPS form, the procedure is quite different: the kick acts on the state in configuration space, whereas the MPS is represented in a space which is a mixture of configuration space and Fock basis. Additionally, we want all of the particles to have the same initial velocity. The total phase imprinting the initial velocity should include factors for all particles:
$\prod_{n=1}^N\exp(i k_0 x_n) = \exp\left(ik_0\sum_{n} x_n\right),$
where $n$ numbers the particles and $N$ is the total number of particles. The sum inside the exponent may be rewritten using particle occupations at each site, i.e.
  $\sum_{n} x_n = \sum_{l} i_l x_l.$
This allows us to use the MPS representation: to kick the MPS, we modify the matrices $\Gamma^{i_l}$: $\Gamma^{i_l} \to \Gamma^{i_l}\cdot e^{i k_0 i_l l  \Delta x}.$ The vectors $\lambda^{[l]}$  are not changed because the kick does not change the properties of the MPS links. The kick preserves the MPS standard form (4) of the many-body state.  

The discretization method was successfully used in a study of a disordered many-body system, where Anderson localization of solitons was observed \cite{Delande2013}. To observe the QBE, the system has to be Anderson localized \cite{Prat2019}. In our work, we use Gaussian uncorrelated disorder:
\begin{equation}
    \overline{V(x)} = 0,\quad \overline{V(x)V(x')} = \gamma\delta(x-x'),
\end{equation}
where $\overline{(\ldots)}$ denotes averaging over disorder realizations and $\gamma$ is the disorder strength. {As explained below, the kicked wavepacket has an energy close to $\hbar^2k_0^2/2m.$ Using the Born approximation, it is easy}  to compute the values of the mean free time and mean free path {at this energy}:
\begin{equation}\label{eq:born_tau_ell}
    \tau_0 = \frac{\hbar^3 k_0}{2m\gamma},\quad \ell_0 = \frac{\hbar^4 k_0^2}{2m^2\gamma}.
\end{equation}

Finally, to observe the QBE, we study the center of mass time evolution: it is evaluated using the average particle density $\overline{n(x,t)}$:
\begin{equation}
    \langle x(t)\rangle = \sum_l x_l \overline{n(x_l,t)}.
\end{equation}
The particle density can be fairly easily {computed}  from the MPS representation: {the} occupation {$n(x_l)$} on {the} $l$-th site depends only on $\lambda^{[l-1]}$, $\lambda^{[l]}$, and $\Gamma^{i_l}$ that are known at each time step.

\section{Weakly interacting bosons}\label{sec:bosons}
\begin{figure}
  \centering
  \includegraphics{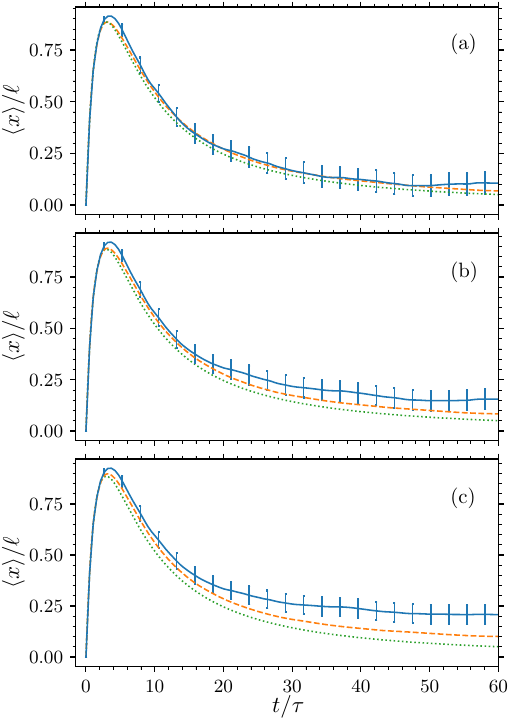}
  \caption{Comparison of results obtained in the many-body simulations (solid lines with error bars), the mean-field simulations (orange dashed lines {with tiny error bars, not shown}), and the single-particle theoretical prediction (green dotted line). Panels correspond to (a) $U = 0.1$, (b) $ U=0.15$, and (c) $U=0.2$. While for the lowest interaction between particles the curves seem to agree (panel (a)), with the increase of the interactions the many-body result saturates at a significantly higher value. 
Mean-field simulations are averaged over $10^5$ disorder realizations.
  }
  \label{fig:tebd_vs_gpe}
\end{figure}
Let us commence our study with the weakly interacting bosons case. 
 {This will allow us to compare our ``quasi-exact'' simulations with results obtained within the mean field approximation that revealed an only partial QBE in the presence of interactions.

The initial state of the system in the mean-field study was a Gaussian wave packet. To mimic this scenario, we prepare the initial state as the ground state of $N$ non-interacting particles in a harmonic trap. Application of an imaginary time evolution {using the} TEBD {algorithm} allows us to prepare the initial state in the MPS form. The frequency of the trap is chosen to match the desired particle density width $\sigma$. Then, the kick with initial momentum $k_0$ is applied to the initial MPS as explained above. The initial particle density is a Gaussian with width $\sigma$: $n_0(x)=N/\sqrt{\sigma^2 \pi}e^{-x^2/\sigma^2}$. In numerical simulations, we use $\sigma=10/k_0 = 2\ell_0$. {Because $k_0\sigma\gg 1$, the wavepacket is quasi-monochromatic with an energy distribution sharply peaked near the energy $\hbar^2k_0^2/2m.$}

In the numerical simulations, we use $1/k_0$ as the unit of length. The system size is $L_\text{size} = 400/k_0$ divided into $L=2000$ lattice sites meaning that $\Delta x = 0.2/k_0$.  We use $N=5$ particles in our simulations. Unfortunately, higher numbers of particles would demand too large  computer resources. The disorder strength is chosen to be $\gamma = 0.1 \hbar^4 k_0^3/m^2$ meaning $k_0\ell_0 = 5$, so we can assume a weak disorder case.  The maximal time of simulations was chosen to be $t_\text{max} = 60\tau_0$. For each interaction strength we used 500 disorder realizations (otherwise stated in figures' captions).

For comparison we present simultaneously the {full many-body results together with}  results obtained within the mean-field approach, using the Gross-Pitaevskii equation \cite{Pethick2008,Pitaevskii2016}:
\begin{equation}\label{eq:gpe}
\begin{split}
    &i\hbar\partial_t \psi(x,t) =\\
    &\left(-\frac{\hbar^2}{2m}\Delta + V(x) + U (N-1)|\psi(x,t)|^2\right)\psi(x,t).
\end{split}
\end{equation}
Since we consider a very small number of particles, the usual factor $g$ multiplying the density part is taken in its exact form  $g=U(N-1)$.

In Fig.~\ref{fig:tebd_vs_gpe}, we show comparisons of the results for many-body and mean-field systems for different interaction strengths. To account for differences between exact scattering mean free time $\tau$ (scattering mean free path $\ell$) and mean free time $\tau_0$ (mean free path $\ell_0$), Eq.~(\ref{eq:born_tau_ell}), we fit the theoretical prediction for CM time dependence \cite{Prat2019} to results obtained for non-interacting particles. This yields $\tau = 0.94\tau_0$ and $\ell = 1.07\ell_0$, which is in a full agreement with expected corrections to the Born approximation{, which are of order $1/k_0\ell_0$ \cite{Akkermans2007book}.}  

Unsurprisingly, we observe that, for nonzero interactions, the boomerang effect is only partial. After the initial ballistic-like motion and the U-turn, typical for the boomerang effect, the CM does not return to the origin but saturates at some finite, interaction strength dependent, position. This closely resembles the behavior observed in the mean-field study \cite{Janarek2020a}.

On the one side, for the lowest presented value of interaction strength $U=0.1$, the many-body and the mean-field solutions are in agreement (within error bars). On the other side, when the interaction strength is higher, for example, in panels (b) and (c) {of} Fig.~\ref{fig:tebd_vs_gpe}, the curves seem to separate and the many-body CM $\langle x(t)\rangle$ saturates significantly higher than the mean-field one.

The interactions present in the system may be understood as a source of dephasing mechanism which destroys Anderson localization, hence the boomerang. From this perspective, it should be natural that, when we treat the interactions without approximation, their impact should be larger, destroying {the} QBE {more efficiently}. However, the simulations include only few particles, we {expect on the general grounds that}  in the limit of a large number of particles, the difference between full quantum and mean field results vanishes.

To {further} study this  difference, we also analyze the one-body reduced density matrix $\rho(x,x') = \langle \hat{\Psi}^\dagger(x')\hat{\Psi}(x)\rangle$, which may be used to analyze correlations in many-body systems, see \cite{Pethick2008}. For this study, we use interaction strength $U=0.2$ with increased disorder strength, so that $k_0\ell_0 = 2.5$. The maximal time of simulation is set to $t_\text{max} = 120 \tau_0$, what should reflect better the long-time limit. 

To quantitatively check the amount of the condensate fraction in the final density matrix $\rho_f(x,x')$, we compute its eigenvalues. The largest eigenvalue represents the condensate fraction \cite{Penrose1956, Yang1962}. For a non-interacting fully condensed system, there is only one eigenvalue $\lambda_0 = N$. When non-zero interactions are present in the system it is no longer true. This approach may be generalized: the interactions decrease the value of $\lambda_0$, however the sum of all eigenvalues is given by the total number of particles, i.e. $\sum_j\lambda_j = N$. The state may be considered a condensate as long as $\lambda_0\sim N$.

In our study, we compute  {the} four largest eigenvalues of the final one-body density matrix. We find the averages {of them} to be $\overline{\lambda_0}/N = 0.147\pm0.028$, $\overline{\lambda_1}/N = 0.110\pm0.016$, $\overline{\lambda_2}/N = 0.083\pm0.011$, $\overline{\lambda_3}/N = 0.064\pm0.007$. {The values for single realizations do not differ much from the averages, the distributions of $\lambda_j$ are narrow.} This clearly indicates that the final state of the system is very far from a true condensate. The GPE describes only the condensate part of the system, while our state consists mainly of particles outside the condensate. Thus, the full dynamics of the system cannot be described with {the} GPE. This fact reinforces the conclusion that the difference between many-body and the mean-field results comes from truly many-body effects.

\subsection{Break time analysis}
\begin{figure}
  \centering
  \includegraphics{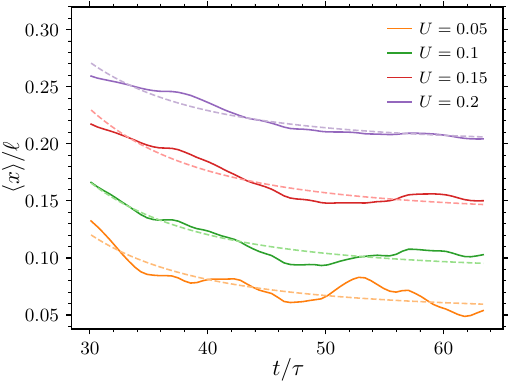}
  \caption{Time evolution of the center of mass (solid lines) in the interval $[30\tau, 64\tau]$ where a fitting of the algebraic decay (dashed lines), Eq.~(\ref{eq:fit}), is performed. When the exponent $\alpha=3$ is used, the resulting fits yield very good results. The $U$ values increase from the bottom to the top as indicated in the legend.}
  \label{fig:fit_algebraic_decay_bosons}
\end{figure}
The destruction of the QBE in the mean-field approximation was successfully described using {the} so-called break time \cite{Janarek2020a}. It is the time {$t_b$ for} which the CM position  {in the} interaction-free case reaches the long-time limit obtained in the presence of  interactions
\begin{equation}\label{eq:break_time_bosons}
    \langle x(t_b) \rangle_{U=0} = \langle x\rangle_\infty.    \end{equation}}
To examine the break time, it is necessary to compute the infinite-time CM position:  $\langle x \rangle_\infty = \langle x (t\to\infty)\rangle$. For  the mean-field approximation, the infinite-time CM position was approximated with the long time average:
\begin{equation}\label{eq:long-time_average}
    \langle x \rangle_\infty = \frac{1}{t_2-t_1} \int_{t_1}^{t_2} \langle x(t)\rangle \diff{x}.
\end{equation}
This was reasonable for large maximal times of numerical simulations, extending up to $2500\tau_0$. In the present study where $t_\text{max}\approx 64\tau,$ such {a} long-time average is not {available}. To overcome this problem, we fit an algebraic decay to the data:
\begin{equation}\label{eq:fit}
    \langle x(t)\rangle = \langle x\rangle_\infty + \frac{\beta}{t^\alpha},
\end{equation}
where $\langle x \rangle_\infty$ and $\beta$ are fitting parameters. The fit is performed in the time interval $[30\tau, t_\text{max}\!\approx\! 64\tau]$. Knowing that, in the non-interacting case, the long-time time dependence is $\langle x (t) \rangle \approx 64\ell \log(t/4\tau) \tau^2/t^2$ \cite{Prat2019}, for the non-interacting case we expect that $\alpha=2$ will return $\langle x\rangle_\infty = 0$, what is confirmed using our numerical data.  For the interacting cases, we find {a} slightly faster decay, thus we use $\alpha=3$ as the exponent in  Eq.~(\ref{eq:fit}).  Figure~\ref{fig:fit_algebraic_decay_bosons} shows a comparison of the numerical data with fitted functions. The fits show very good agreement with the data. It also turns out that the overall fitting results, i.e. the values of $\langle x \rangle_\infty$, only slightly depend on the exponent value $\alpha$ in Eq.~(\ref{eq:fit}) and the time fitting interval.

\begin{figure}
  \centering
  \includegraphics{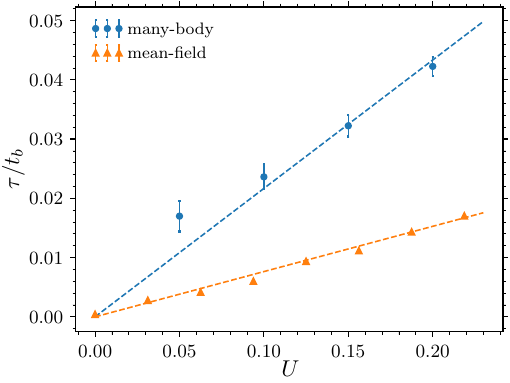}  
  \caption{Inverse of the break time $t_b$ computed for the many-body simulations (blue points with error bars, computed from the fits of the algebraic decay, Eq.~(\ref{eq:fit})) and the mean-field simulations (orange triangles, calculated using long-time averaging, Eq.~(\ref{eq:long-time_average})) versus the interaction strength $U$.  Dashed lines present the best linear fits $\tau/t_b=aU$, with slope coefficients $a_\text{many-body}=0.22$ and $a_\text{mean-field}=0.076$. The mean-field data is clearly linear, as expected. For the many-body results, with a small deviation of the point for $U=0.05$, the points strongly suggest linear dependence. The error bars represent the {uncertainty on} the break time based on the {error bars for} the final center of mass position value.}
  \label{fig:one_over_tb_vs_v_bosons}
\end{figure}

Having the estimate for $\langle x\rangle_\infty$, we
can find the break time with the help of Eq.~\eqref{eq:break_time_bosons}.
Fitting errors on the infinite time CM position $\langle x \rangle_\infty$ allows us to calculate the error bars on the break times for various interaction strengths, $U$.
In  analogy with the mean-field study \cite{Janarek2020a}, we expect the inverse of $t_b$ to be proportional to $U$, a measure of the interaction energy in the system. 

The dependence of $1/t_b$ versus $U$ is shown in Fig.~\ref{fig:one_over_tb_vs_v_bosons}, where we present results for the many-body and the mean-field simulations (where $\langle x\rangle_\infty$ is computed from the long-time average, Eq.~(\ref{eq:long-time_average})). While for the mean-field results the dependence is obviously linear, the many-body result also suggests a linear behavior, with a small deviation of the point with $U=0.05$. This point, the lowest value of the interaction strength, requires the longest time of evolution to saturate around the true $\langle x \rangle_\infty$ value. The corresponding infinite time CM position value may be overestimated, {in turn} underestimating the break time. On the opposite side, for stronger interactions, the linearity is better. This is also related to the fact that the final infinite time CM position values are higher, hence, to compute the break time, a shorter time evolution is sufficient.

The fact that the break time is much shorter for the full many-body calculation than in the mean-field approximation, emphasizes the importance of quantum fluctuations. This is also supported by the analysis of the average one-body density matrix presented above.

Before moving to the next part of our study, we should make a comment on the many-body localization phenomenon (MBL), which may be present in many-body interacting disordered systems (for reviews see \cite{Alet2018, Abanin2019}). Although, we study a disordered many-body system, we are not in the MBL regime. Typically, MBL is studied in systems with much higher interaction strengths (for example of bosonic systems see  \cite{Sierant2017, Sierant2018}) than considered in our work, where $U_0/J_0\ll 1$ (translating $\Delta x$, and $U$ to Bose-Hubbard model parameters). The other important factor is the density of particles -- the average filling in our system, taking into account only the sites occupied by the initial density profile, is very low, $n\approx0.1$. Together with the small number of particles considered, this does not allow for a comparison with other studies of interacting bosons on a lattice. Finally, the disorder strength used in our study corresponds to Anderson localization {in the} weak disorder regime, which should not be sufficient to induce many-body localization effects.

\section{Strongly interacting bosons: the Tonks-Girardeau limit}\label{sec:tg}
Let us now consider a second entirely different situation - the case of very strong interactions.
A one-dimensional system of bosons with repulsive contact interactions may be described by the Lieb-Liniger model \cite{Lieb1963}:
\begin{equation}\label{eq:lieb_liniger}
  H =  \sum_{j=1}^N \left( -\frac{\hbar^2}{2m}\frac{\partial^2}{\partial x_j^2} + V(x_j)\right) + U\sum_{1\leq j < k\leq N} \delta(x_j - x_k),
\end{equation}
where $U>0$ is the coupling constant and $m$ denotes the atom mass. The model is frequently characterized by a dimensionless parameter $\zeta = mU/\hbar^2n$, where $n=N/L$ being the average density of bosons, and $L$ is the system length. When
 $\zeta=0$, the model corresponds to free bosons while $\zeta\to\infty$ is called the Tonks-Girardeau limit.

The Tonks-Girardeau (TG) gas describes impenetrable (or \emph{hard-core}) bosons, which can be  mapped to non-interacting spinless (spin-polarized) fermions \cite{Girardeau1960, Girardeau2000}. The model can be solved
exactly in the free case $V=0$ (for details see \cite{Cazalilla2011}). Reference~\cite{Olshanii1998} showed that the Tonks-Girardeau gas can be obtained in cold atom experiments, and the experimental observations of hard-core Rubidium bosons were reported shortly after in \cite{Paredes2004,Kinoshita2004}.

Even though {the} TG gas is highly correlated, Anderson localization is not destroyed by the interactions. TG particles map to non-interacting fermions, hence Anderson localization is present in the system: non-interacting fermions are fully localized in a one-dimensional system. Anderson localization of the TG gas was discussed in \cite{Radic2010}.

In an Anderson localized system, we expect to observe the full quantum boomerang effect for particles with non-zero initial velocity. We perform numerical simulations to study the center of mass temporal evolution, using the same methods as in the case of weakly interacting bosons. In order to simulate {the} TG gas, we use a trick: the MPS representation has a parameter  $i_\text{max}${, the maximum number of bosons on a given site (the local Hilbert space thus has dimension  $i_\text{max}+1$)}. For $N$ bosons, it is natural that $i_\text{max}\approx N$ which allows the MPS to represent faithfully states with many particles at one site. This parameter can be used in the other way: we restrict the number of particles occupying one site by setting $i_\text{max}=1$, effectively realizing the concept of impenetrability of the Tonks particles. {Note that the local Hilbert space has dimension 2, explaining why it can be mapped on spinless fermions, where the local Hilbert space is spanned by states with 0 or 1 fermion.}

On the numerical side, our results have been simulated in a similar way to the weakly interacting bosons. The main difference is that, in the Tonks-Girardeau gas, we enlarged the discretization constant, so that $\Delta x = 1/k_0$. By using larger $\Delta x$, we can decrease the number of lattice sites  in the simulations to $L=500$ and scale down {the} CPU-time. The main effect of larger $\Delta x$ is its influence on the dispersion relation. As we show in the next sections, apart from the change of velocity due to not ideal discretization, the quantum return to the origin still can be analyzed.

As opposed to above studies of QBE, here we cannot use {a} Gaussian wave packet as the initial state of the system. It is very different from the ground state of TG particles in a harmonic trap.  Nevertheless, the ground state of TG particles in a trap can be computed. Due to {the} mapping to fermions, it can be easily found in the absence of the disordered potential. Fermions cannot occupy the same eigenstate of the system, hence the state with the lowest energy has {the} following structure in the Fock basis (ordered by increasing energy): $\ket{\text{GS}} = \ket{11\ldots10\ldots}$, with $N$ particles occupying {the} $N$ {single particle} states with the lowest energy. Then, the particle density can be calculated in a straightforward way:
\begin{equation}\label{eq:tg_groundstate}
    n_\text{TG}(x) = \sum_{n=0}^{N-1}|\psi_n(x)|^2,
\end{equation}
where $\psi_n(x)$ denotes a single-particle eigenstate of the trap. The density is much broader than the harmonic oscillator's ground state for a single particle. On the numerical side, the initial state is prepared using the imaginary-time evolution {in presence of interactions, but in the absence of disorder} followed by a velocity kick, similarly to the weakly interacting bosons case.

\begin{figure}
  \centering
  \includegraphics{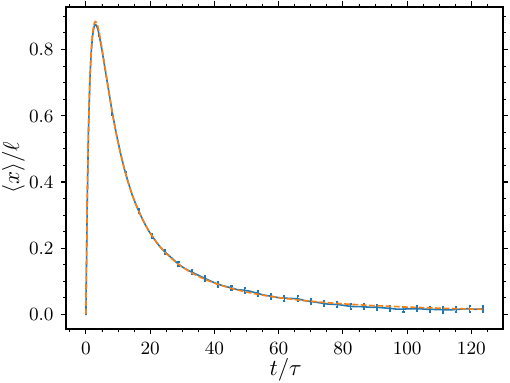}
  \caption{Time evolution of {the} center of mass $\langle x \rangle$ for {the} Tonks-Girardeau gas (solid blue line with error bars) compared with {the} single-particle theoretical prediction (orange dashed line) \cite{Prat2019}. {The r}esult is fitted using {the} theoretical boomerang prediction to {adjust} the mean scattering time $\tau$ and length $\ell$. The numerical data perfectly agree with the theoretical curve. The results have been averaged over 10000 disorder realizations. Error bars represent statistical average uncertainties.}
  \label{fig:tg_boomerang}
\end{figure}

\begin{figure}
  \centering
  \includegraphics[scale=1]{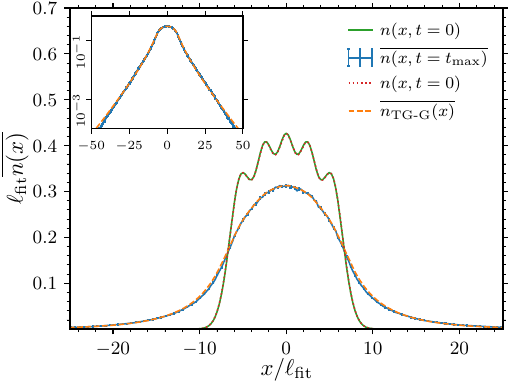}
  \caption{{Initial (green solid line) and } final density profile for kicked hard core bosons (blue solid line with error bars), compared with {the theoretical initial particle density (red dotted line)}, Eq.~(\ref{eq:tg_groundstate}), and the Tonks-Girardeau-Gogolin profile (orange dashed line){, Eq.~(\ref{eq:tgg_profile})}. {The numerical data for the initial and final times agree fully with the initial density and with the theoretical Tonks-Girardeau-Gogolin profile, respectively.}  The inset shows the theoretical and numerical final profiles to show agreement even in the exponentially decaying tails.  }
  \label{fig:tg_final_density}
\end{figure}

Figure~\ref{fig:tg_boomerang} presents the time evolution of the CM $\langle x(t)\rangle$ for the  {TG} gas. 
It faithfully follows the single-particle QBE. To show the agreement between the numerical data and the theoretical prediction \cite{Prat2019}, we perform a fitting procedure which accounts for the difference between the exact mean free time $\tau$ (mean free path $\ell$) and the mean free time $\tau_0$ (mean free  path $\ell_0$) computed using the Born approximation, Eq.~(\ref{eq:born_tau_ell}). 

After such a fit, the agreement between the TG gas and the theoretical prediction is excellent. The disorder strength used should result in $k_0\ell_0 = 5$. The fitted exact values of mean free time and path are $\tau=0.97\tau_0$ and $\ell=0.9\ell_0$, 
being thus consistent with the Born approximation.

There is, however, a slight caveat. The particles of the {TG} gas have slightly different energies, because they correspond to different eigenstates of the harmonic potential. This should mean that each particle has a different mean free time, hence $\langle x (t)\rangle$ should be a superposition of the boomerang curves with different $\tau$.
The energy of the $n$-th eigenstate of the harmonic potential is $(n + \frac{1}{2})\hbar\omega$, where $\omega$ is the frequency of the {harmonic oscillator}. In our analysis we use kicked states, and the kick adds $\hbar^2k_0^2/2m$ to the total energy. If $\hbar^2k_0^2/2m \gg (n + \frac{1}{2})\hbar\omega$, we may assume that all states have roughly the same scattering  mean free time and path. This is the case in our simulations, where $\omega=0.01$.
The small dispersion of energies does not influence the final $\langle x(t)\rangle$, and we observe the universal boomerang curve.

We also study the final particle density. It is symmetric and has exponentially decaying tails. Although \cite{Radic2010} used a slightly different initial state (ground state of the trap including the disorder), a similar behavior of the tails in their simulations was reported. After our observation that the boomerang effect is described by a single-particle theoretical result, we construct an infinite-time density profile based on the (single-particle) Gogolin profile \cite{Gogolin1976}:
\begin{equation}\label{eq:gogolin}
\begin{split}
    &\overline{|\psi^\text{Gogolin}_\ell(x,t=\infty)|^2} = \\
    &\int_0^\infty \frac{\diff{\eta} \pi^2}{32\ell} \frac{\eta(1+\eta^2)^2 \sinh(\pi\eta)e^{-(1+\eta^2)|x|/8\ell}}{(1+\cosh(\pi\eta))^2},
\end{split}
\end{equation}
which depends on the mean free path $\ell$. 
{As explained in \cite{Gogolin1976}, this density profile is the theoretical prediction at infinite time for a single particle initially located at $x=0$ and evolving in the presence of a disordered potential. In our case, t}he final density should be given by the convolution of the Gogolin profile with the initial particle density $n_\text{TG}(x)${, Eq.~(\ref{eq:tg_groundstate})}:
\begin{equation}\label{eq:tgg_profile}
  \overline{n_\text{TG-G}(x)} = \int_{-\infty}^{+\infty}\diff{x'} n_\text{TG}(x-x') \overline{|\psi_\ell^\text{Gogolin}(x')|^2}.
\end{equation}

In the analysis of the final density profile, we also fit $\overline{n_\text{TG-G}(x)}$ to numerical data. The numerical calculation of the Tonks-Girardeau-Gogolin profile for $x/\ell \gg 1$ is laborious, thus we fit the profile only around $x=0$ for several points. The value of the fitted mean free path is $\ell_\text{fit}\approx4.025/k_0$. The mean free path extracted from the center of mass time evolution is $\ell=4.5/k_0$. Taking into account the fact that $\ell_0 k_0=5$, so that corrections to the Born approximation may be visible, the agreement between $\ell_\text{fit}$ and $\ell$ is good.

Figure~\ref{fig:tg_final_density} shows both the numerical and fitted final densities as well as the initial density profile. Our crude fitting method gives neverthless very good results -- the numerical and theoretical infinite-time densities agree {perfectly}. The inset presenting the densities in a logarithmic scale shows almost no difference also in the wings, far from the region of the fit.

\section{Strongly interacting bosons: mapping to weakly interacting fermions}\label{sec:fermions}
For an arbitrary interaction strength in Hamiltonian~(\ref{eq:lieb_liniger}), the bosonic model can be mapped to interacting fermions~\cite{Sen1999, Cheon1999, Sen2003}. The interaction is much more complicated, it is mapped to a momentum-dependent attractive interaction~\cite{Grosse2004}. Fermions are governed by the following Hamiltonian:
\begin{equation}\label{eq:fermionic}
  H_F = \sum_{j=1}^N \left(-\frac{\hbar^2}{2m}\frac{\partial}{\partial x_j}  + V_\text{ext}(x_j)\right) + V_F,
\end{equation}
where $V_F$ denotes the fermionic interaction term:
\begin{equation}\label{eq:fermionic_interaction}
  V_F = \frac{\hbar^4}{m^2 U} \!\!\sum_{1\leq j<k\leq N} \!\! \left(\partial_{x_j} - \partial_{x_k}\right) \delta(x_j - x_k) \left(\partial_{x_j} - \partial_{x_k}\right).
\end{equation}
The eigenfunctions of the Lieb-Liniger Hamiltonian~(\ref{eq:lieb_liniger}) coincide with the eigenstates of Hamiltonian~(\ref{eq:fermionic}) when particle coordinates $x_j$ are ordered and their sign is changed upon exchange of the particle coordinates. The models have the same eigenspectra. The fermionic interaction strength is proportional to $U^{-1}$, see Eq.~(\ref{eq:fermionic_interaction}). In order to simplify the notation, in the following sections we use $U_F=U^{-1}$ to represent the interaction strength between the fermions.

The mapping can be used to study systems in different potentials including disordered ones, e.g. the fluid-insulator transition for strongly interacting bosons was studied in~\cite{Michal2016}.

Above, we argued that the simulations of disordered many-body systems require large amounts of computational resources. To compute simulations of strongly interacting bosons, we allow at most two particles at one site, $i_\text{max}=2$. In the case of weak interactions, such constraint would  change the results and simulations would not be faithful. On the other hand, when the interactions are strong, the probability of having more than two particles at one site is small being energetically very costly \cite{footnote1}.
Additionally, we keep $\Delta x =1/k_0$ as in the TG gas case. Altogether that allows us to save computational resources and calculate the {temporal} evolution for longer times than for weakly interacting bosons. Let us stress that here we cannot be guided by the mean-field analysis.

\begin{figure}
  \centering
  \includegraphics{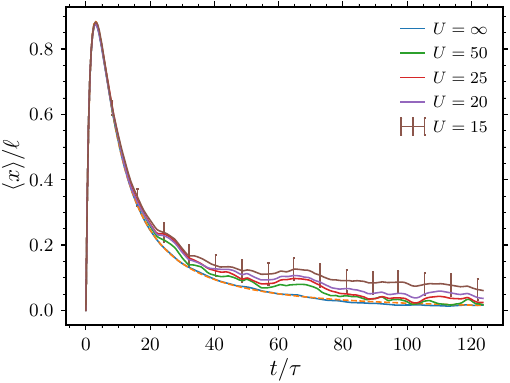}
  \caption{{Temporal evolution of the center of mass position} for different values of {interaction strength $U$, decreasing from the bottom to the top, in the strong interaction limit}.  Similarly to the mean-field case and weakly interacting bosons, the short time evolution is almost unaffected by interactions. At longer times, the center of mass {saturates} at finite values. Error bars indicate statistical  errors and are shown only for one curve to indicate their magnitude. Orange dashed curve shows the theoretical {center of mass} {temporal} evolution (cf. Fig~\ref{fig:tg_boomerang}). }
  \label{fig:tg_cmp_interactions}
\end{figure}
At the qualitative level, the effect of interactions on the QBE \new{is likely to} not depend on their details. For strongly interacting bosons  (weakly interacting fermions), we also expect that interactions will weaken Anderson localization. \new{Any interaction has a characteristic energy scale which
should translate into some break time, beyond which the QBE should be
broken.} The interactions, which are considered as an effective dephasing mechanism, lead to the destruction of coherence between scattering paths, and finally to destruction of the full QBE.

Figure~\ref{fig:tg_cmp_interactions} presents the result of the CM  time evolution. Similarly to the non-interacting case, after the initial ballistic evolution, the CM is reflected towards the origin. Analogously to the mean-field and weakly interacting bosonic cases, the destruction of the boomerang effect is visible in the long time regime. For all situations with finite $U$ (nonzero effective interaction between fermions $U_F$), we observe that the return is not complete: the infinite time CM position saturates at some nonzero value. The figure shows also the statistical error bars. Because the number of disorder realizations is small, the errors are relatively large. Nonetheless, the effect of interactions is clearly visible and can be analyzed taking into account the uncertainties. The limited maximal time of evolution does not allow us to study in detail the mean square displacement of the particle density.

The main observation is that the boomerang effect is only partial, even though the effective interactions between fermions are attractive and fairly complicated. There is no qualitative difference between the results of the mean-field approximation, weakly interacting bosons and weakly interacting fermions. Interactions weaken the QBE.

\begin{figure}
  \centering
  \includegraphics{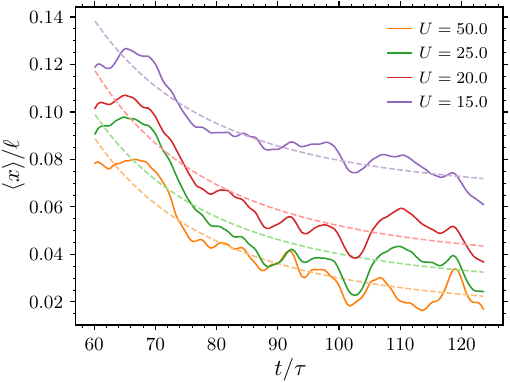}
  \caption{{Temporal} evolution of the {center of mass position for strongly interacting bosons} (solid lines) and fits of the algebraic decay, Eq.~(\ref{eq:fit}) (dashed lines). As indicated in the figure the values of $U$ decrease from bottom to top curves. Similarly to Fig.~\ref{fig:fit_algebraic_decay_bosons}, with the exponent $\alpha=3$, the resulting fits yield satisfactory results.}
  \label{fig:fit_algebraic_decay}
\end{figure}

As for weakly interacting bosons, we analyze the final CM position. 
We use, as before,  the algebraic fit{, Eq.~(\ref{eq:fit})}  to extract the infinite time CM position {from data}
 in the time interval $[60\tau,120\tau]$.
 As for weakly interacting bosons we assume $\alpha=3$ for the fits. Figure~\ref{fig:fit_algebraic_decay} shows a comparison of the numerical data with fitted functions. The data show high correlation between different interaction strengths because we use the same disorder realizations. Also in this case, we have checked that the overall fitting result is almost independent of the exponent value $\alpha$ in Eq.~(\ref{eq:fit}).

\begin{figure}
  \centering
  \includegraphics{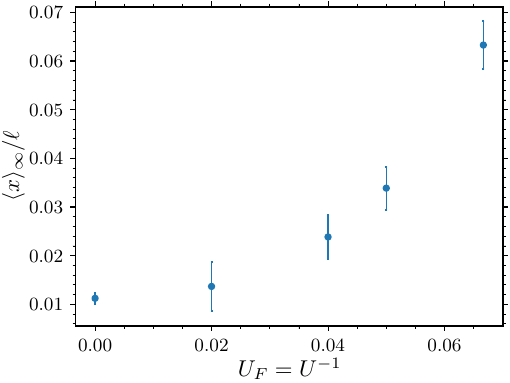}
  \caption{Final center of mass position $\langle x \rangle_\infty$ versus $U_F=U^{-1}$. The errors for the points result from the fitting of the decay Eq.~(\ref{eq:fit}). Like in the mean-field study, the dependence of the final center of mass position on the effective interaction strength between fermions $U_F=U^{-1}$ is quadratic.}
  \label{fig:final_cmp_vs_u}
\end{figure}
In Fig.~\ref{fig:final_cmp_vs_u}, we present the dependence of  $\langle x\rangle_\infty$ on the effective interaction strength $U_F$ between fermions. As in the case of the mean-field approximation \cite{Janarek2020a}, for the smallest values of the interaction strength, the dependence seems to be quadratic. This confirms that the observed breakdown of the QBE does not depend on the details of the interactions present in the system.

\begin{figure}
  \centering
  \includegraphics{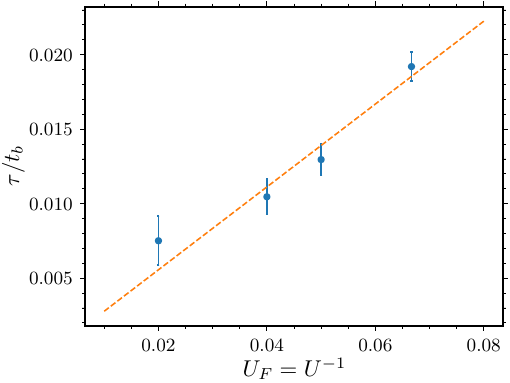}
  \caption{Inverse of the break time $t_b$ versus $U_F=U^{-1}$ calculated for the final center of mass position $\langle x \rangle_\infty$ by fitting an algebraic decay, Eq.~(\ref{eq:fit}). The {error bars} are calculated using the  {uncertainty on}  $\langle x \rangle_\infty$. The data strongly suggest a linear dependence. The dashed line presents the best linear fit $\tau/t_b = 0.28 U_F = 0.28/U$.}
  \label{fig:one_over_tb_vs_u}
\end{figure}
Given the results presented in the previous section, we may ask whether the destruction of the boomerang effect for strongly interacting bosons can be effectively described using the break time, a universal parameter used to capture the influence of the interactions. 

\subsection{Break time -- boomerang effect}

For the weakly interacting bosons, the use of break time was a natural extension of the mean-field approximation. In the case of strongly interacting bosons (mapping to weakly interacting fermions), this has to be analyzed anew. Figure~\ref{fig:final_cmp_vs_u} shows the approximately quadratic dependence of the  $\langle x \rangle_\infty$ on the effective interaction strength between fermions $U_F$. 

Figure~\ref{fig:one_over_tb_vs_u} shows the dependence of the inverse of the break time, $1/t_b$, on the effective interaction $U_F=U^{-1}$ suggesting a linear behavior. Similarly to the weakly interacting bosons case, the point for the weakest interactions slightly deviates from the linear dependence. When the QBE is only moderately affected by the interactions, the time evolution has to be very long to extract the exact value of the infinite time CM position. When $\langle x \rangle_\infty$ is overestimated, the corresponding $t_b$ is smaller than the exact value.

The results are very similar to those obtained for the weakly interacting bosons, see Fig.~\ref{fig:one_over_tb_vs_v_bosons}. It means that the underlying mechanism of the destruction of the QBE is independent of the type of interactions. The destruction of the  QBE may be fully characterized by a single parameter, the break time $t_b$, proportional to the interaction strength between the particles.

{It is possible to understand semi-quantitatively the $1/U$ dependence of the break time. At infinite $U,$ the dynamics of the system takes entirely place in the sub-space spanned by occupation numbers $i=0$ and $i=1$ on each site of the Bose-Hubbard Hamiltonian,  Eq.~(\ref{eq:bose_hubbard}), and one observes full QBE. When $U$ is large, but finite, the state with occupation number $i=2$ also comes into the game. However, due to interaction, its energy is larger by $U,$ while the coupling with $i=0,1$ states is typically of the order of $J$. An example is the coupling between states $|0,2\rangle$ and $|1,1\rangle$ on two neighboring sites. The perturbation brought by $i=2$ states is thus expected, at lowest order, to shift the energy levels in the $i=0,1$ subspace proportionally to $J_0^2/U_0.$ In the absence of this shift, the QBE is full. It is thus reasonable to expect that, for finite $U_0=U/\Delta x,$ it will take a time $\hbar/(J_0^2/U_0)$ before the QBE is affected. In other words, we expect the break time to be roughly $U_0/J_0^2$.}

 It turns out that the boomerang break times (expressed in units of the  scattering mean free time) agree within several percent with  {the rough} estimate:
\begin{equation}
	\begin{aligned}
		&U_0 = 50\quad \frac{U_0}{J_0^2} &=\,\,& 200 \quad &t_b = 133.2\tau,\\
		&U_0 = 25\quad \frac{U_0}{J_0^2} &=\,\,& 100 \quad &t_b = 95.6\tau, \\
		&U_0 = 20\quad \frac{U_0}{J_0^2} &=\,\,&  80\quad &t_b = 77.2\tau, \\
		&U_0 = 15\quad \frac{U_0}{J_0^2} &=\,\,&  60\quad &t_b = 52.1\tau.
	\end{aligned}
\end{equation}
As explained above, the break time for the highest interaction strength $U_0$ is, most probably, underestimated.

\subsection{Break time for the entropy of entanglement}
\begin{figure}
  \centering
  \includegraphics[scale=1.]{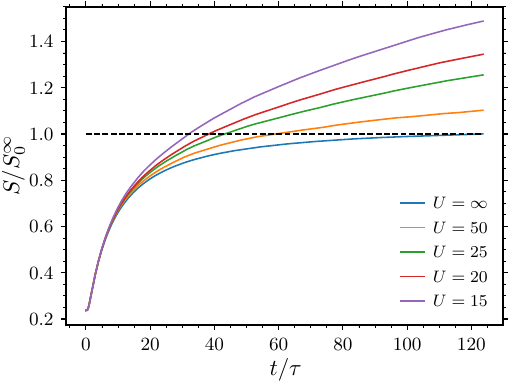}
  \caption{{Temporal} evolution of the entropy of entanglement (average{d} over all possible bipartitions) for different values of the interaction strength $U$, decreasing from the bottom to the top. $S_0^\infty$ denotes the final value of the entropy in the Tonks-Girardeau gas. }
  \label{fig:ent_vs_time}
\end{figure}
In the simulations, we can also observe another interaction-driven phenomenon, which can be characterized by its own time scale. Due to the interactions, we observe a growth of the entropy of entanglement in the system. Does it increase on the same time scale as $t_b$?

Figure~\ref{fig:ent_vs_time} shows the time evolution of the entropy of entanglement computed as an average over {all} possible bipartitions\new{:
\begin{equation}
S=-\frac{1}{L-1} \sum_{i=1}^{L-1} \sum_\alpha  (\lambda^{[i]}_{\alpha})^2 \ln (\lambda^{[i]}_{\alpha})^2,
\end{equation}
-- c.f.  Eq.~(\ref{eq:psi_mps}) -- where different $i$'s in the sum correspond to different bonds of the chain of length $L$.}  For the Tonks-Girardeau gas case, apart from the initial growth, the entropy saturates, what is also confirmed by the analysis of the supremum of the entropy over possible bipartitions {(not shown)}. We denote the final value of the entropy for the Tonks-Girardeau gas by $S_0^\infty$. When the interactions are finite  ($U_F=U^{-1}\neq0$), the entropy grows further.

\begin{figure}
  \centering
  \includegraphics[scale=1.0]{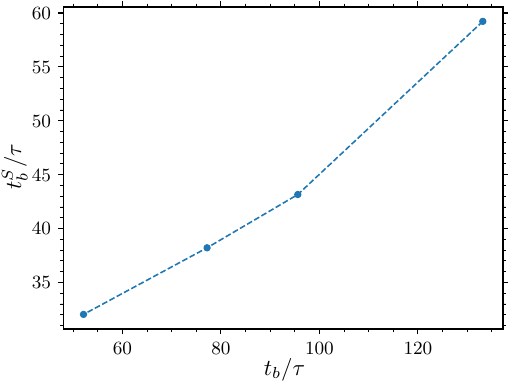}
  \caption{Entropy based break time $t^S_b$ plotted versus {the} boomerang break time $t_b$. The values of break times are comparable within a factor 2. The dependence is more or less linear, the slight deviation for the point around $t_b\approx13
0\tau$ originates probably in the overestimation of  $\langle x \rangle_\infty$ due to too short time evolution.}
  \label{fig:break_times_comparison}
\end{figure}
We can define a characteristic time scale called entropy break time, denoted by $t_b^S$, for which the entropy between the interacting particles exceeds the final value of the Tonks-Girardeau gas entropy $S_0^\infty$. We calculate its value from the following relation:
\begin{equation}
  S(t^S_b)(U) = S_0^\infty,
\end{equation}
where for the left-hand side, we use the data for nonzero interactions. Figure~\ref{fig:break_times_comparison} presents a comparison of the boomerang break time and entropy break time. The relation between the break times is approximately linear{:} $t^S_b/t_b\approx 0.5$.

\section{Conclusions}\label{sec:conclusions}
In this work, we have discussed the effect of interactions on the quantum boomerang effect using {a} quasi-exact many-body approach. On the numerical side, the  simulations have been performed using the time evolving block decimation algorithm based on matrix product states. This has allowed us to study the weakly interacting bosons, the Tonks-Girardeau gas, and strongly interacting bosons which can be mapped to weakly interacting fermions.

The first part of our study has shown that the effect of weak interactions between bosons is qualitatively similar to the behavior in the mean-field approximation \cite{Janarek2020a}. {However, in the present work}, the interactions are not approximated, which strengthens their effect on the destruction of the boomerang effect: the final center of mass positions are higher than in the mean-field approximation. This translates {in}to shorter break times {for} the many-body system. In the simulations, the total number of particles is not very high, so, to support this conclusion, we have also analyzed the features of the average one-body density matrix which have clearly shown that the condensate fraction in our system is very low. Hence, the observed phenomena are necessarily {beyond} the mean-field analysis.

In the second part, we have shown that the particles of the Tonks-Girardeau gas undergo the full boomerang effect. Apart from agreement between the numerical and theoretical results for the center of mass evolution, we have shown that the final particle density is given by {the} convolution of the Gogolin profile and the initial particle density. 

Finally, we have presented that, in the case of {finite} {strong} interactions between bosons (that is effective weak interactions between fermions), the boomerang effect is only partial. To study the destruction of the QBE in detail, we have calculated the break time and shown that is proportional to the interaction strength between bosons, i.e. inversely proportional to the effective interaction strength between {the} fermions. Moreover, from the analysis of the entropy of entanglement, we have computed another characteristic time and shown that this time is comparable and proportional to the break time.

{Altogether, our results strongly suggest that the breaking of the QBE by interactions is a rather simple and universal phenomenon, which can be described by a single parameter, the break time, independently of the details of the interaction and whether a full many-body or a mean-field description is used.}

Possible future studies of the many-body quantum boomerang could include analysis of the phenomenon for a composite particle, i.e. {a} soliton. In~\cite{Delande2013}, many-body Anderson localization of a bright soliton was shown using very similar numerical tools. It would be very interesting to check whether such a composite object undergoes the quantum boomerang effect. 
{While the present work was restricted to rather weak disorder - where analytical predictions for $\langle x(t)\rangle$ are possible - the regime of both strong disorder and strong interactions would be very interesting, especially in the regime of many-body localization.}

\acknowledgments
We thank Nicolas Cherroret for useful discussions.
This research has been supported by the National Science Centre (Poland) under Project No. 2016/21/B/ST2/01086 (J.J.) and  2019/35/B/ST2/00034 (J.Z.). We acknowledge support of the Foundation for Polish Science (FNP)  through the first Polish-French Maria Sklodowska-Pierre Curie reward received by D. Delande and J. Zakrzewski. J.~Janarek  acknowledges also the support of French Embassy in Poland through the \emph{Bourse du Gouvernement Fran\c cais} program. We acknowledge the support of the PL-Grid structure which made numerical calculations possible. 


\begin{thebibliography}{50}
\bibitem [{foo()}]{footnote1}%
  \BibitemOpen
  \href@noop {} {}\bibinfo {note} {Remember that also the total number of
  particles is rather low, $N=5$ and that the interaction energy in the
  Bose-Hubbard model for $n$ particles at the same site is $U n(n-1)/2=3U$ for
  $n=3$.}\BibitemShut {Stop}%
\end{thebibliography}
%

\end{document}